\documentclass{ttp_DSL2006}

\usepackage[dvipsone]{graphicx}
\usepackage[intlimits]{amsmath}
\usepackage{amssymb,gensymb}
\usepackage{exscale}
\usepackage[colorlinks,urlcolor=blue]{hyperref}
\usepackage{booktabs,tabularx}
\newcommand{\hsic}{4H-SiC}
\newcommand{\defZ}{Z${}_\text{1/2}$}
\newcommand{\defEH}{EH${}_\text{6/7}$}
\newcommand{\defEI}{EI${}_\text{5/6}$}
\newcommand{\EC}{E${}_\text{C}$}
\newcommand{\EV}{E${}_\text{V}$}
\newcommand{\VC}{V${}_\text{C}$}

\newcommand{\VSi}{V${}_\text{Si}$}
\newcommand{\VSimmm}{V${}_{\text{Si}}^{3-}$}
\begin{document}

\title{Large-scale electronic structure calculations of vacancies in 4H-SiC using the Heyd-Scuseria-Ernzerhof screened hybrid density functional} 

\author{Tam\'{a}s Hornos \inst{1}$^{,\rm{a}}$ \textbf{,} \'{A}d\'{a}m Gali \inst{2}$^{,\rm{b}}$ and Bengt G. Svensson \inst{1}$^{,\rm{c}}$}

\institute{Department of Physics, University of Oslo, Sem S\ae landsvei 24, N-0316, Oslo, Norway
\and
Research Institute for Solid State Physics and Optics, Budapest, P.O.B. 49, H-1525, Hungary
}
\maketitle

\vspace{-3mm}
\sffamily
\begin{center}
$^{a}$tom.hornos@gmail.com, $^{ b}$agali@eik.bme.hu, $^{c}$b.g.svensson@fys.uio.no
\end{center}

\vspace{2mm} \hspace{-7.7mm} \normalsize \textbf{Keywords:} sic, carbon vacancy, silicon vacancy, ab-initio calculation, hybrid functional\\

\vspace{-2mm} \hspace{-7.7mm}
\rmfamily
\noindent \textbf{Abstract.} Large-scale and gap error free calculations of the electronic structure of vacancies in \hsic~have been carried out using a hybrid density functional (HSE06) and an accurate charge correction scheme. Based on the results the carbon vacancy is proposed to be responsible for the \defZ~and \defEH~DLTS centers.

\vspace{-2mm}
\section{Introduction}
\vspace {-3mm}
The potentials of SiC are compromised by the presence of persistent deep-level defects. Two of the most prominent ones are the \defZ~\cite{PhysRevB.58.R10119} and the \defEH~centers \cite{hemmingsson:6155}. The origin of these centers is not well-established but they are both intrinsic-related defects of fundamental nature which appear also in as-grown material. Possible candidates are carbon vacancy related defects and small carbon clusters at the silicon site. DLTS studies on electron-irradiated \hsic~samples found that irradiation greatly increases the defect concentrations and the centers persist in abundant quantity up to high temperatures. These defects strongly influence charge carrier lifetimes which are crucial for bipolar SiC device performance. In particular, due to its negative-$U$ property, the \defZ~defect is a very efficient electron trap since it can capture two electrons.

Previous \textit{ab initio} calculations proposed several microscopic models for the \defZ~center. Eberlein \textit{et al.} suggested a nitrogen-carbon interstitial model. \cite{PhysRevLett.90.225502} However, later it was shown experimentally that this center does not contain nitrogen. \cite{L.StorastaAnneHenryJ.PederBergman2004} Zywietz \textit{et al.} \cite{PhysRevB.59.15166} and Bockstedte \textit{et al.} \cite{PhysRevLett.105.026401} reported negative-$U$ property for the carbon vacancy (\VC) which might be the \defZ~defect. Gali \textit{et al.} \cite{PhysRevB.73.033204} studied carbon-interstitials and proposed a 3-carbon structure showing negative-$U$ behavior as a candidate for the \defZ~center. Usually, the \defZ~and \defEH~centers appear together, this holds especially for as-grown samples and also after irradiation with low energy ($\lesssim$400 keV) electrons \cite{danno:113728} while after irradiation with MeV electrons or ions the concentrations of the two defects deviate from a one-to-one correspondence \cite{alfieri:043518}. In the latter case, the \defEH~defect level is found to be broad with overlapping contributions while in the former case one single contribution dominates \cite{danno:113728,wong-leung:142105}. Based on these results, Danno \textit{et al.} \cite{danno:113728} suggested that the \defZ~and \defEH~levels stem from the same microscopic origin, namely the carbon vacancy.

Former \textit{ab initio} simulations on intrinsic defects in SiC were done by standard DFT in relatively small supercells with \textit{ad-hoc} charge corrections, except the study done by Gali \textit{et al.} \cite{PhysRevB.73.033204} and Bockstedte \textit{et al.} \cite{PhysRevLett.105.026401} who utilized a non-selfconsistent method to correct the band gap error. Even so, the lack of highly accurate calculations prevented unambiguous identification of \defZ~and \defEH~defects so far. In this study we have investigated the fundamental monovacancies in 4H-SiC and focused on accurate calculation of their occupation levels.

\section{Method}
\vspace {-3mm}
It is well-known that standard DFT suffers from band gap error which affects energy position of defect levels as well. A highly accurate study needs large supercells and a methodology which removes the band gap error as well as corrects charged supercells. We carried out large-scale \textit{ab initio} calculations of silicon and carbon vacancies in \hsic, where the LDA band gap error was corrected employing the Heyd-Scuseria-Ernzerhof hybrid functional (HSE06). The HSE06 functional has been proven to successfully reproduce energy levels for a wide range of defects in group-IV semiconductors \cite{PhysRevB.81.153203}, and the calculated band gap of 4H-SiC of 3.16 eV compares well with the experimental value of 3.27 eV.

First, geometries were optimized by standard DFT in a 576-atom supercell, then the geometries were recalculated with the hybrid functional without further relaxation. Calculations were performed using the plane-wave projector augmented-wave (PAW) method with Perdew-Burke-Ernzerhof (PBE) GGA functional and HSE06 hybrid functional as implemented in the VASP code. \cite{paier:154709} The plane-wave cutoff for the wavefunction expansion (and for that of the charge density􏰉) was set to 420 eV (840 eV). For k-point sampling only the $\mathbf{\Gamma}$-point was used. Besides the band gap error, charged supercells imply another challenge, and we utilized the correction scheme suggested by Lany \textit{et al.} \cite{PhysRevB.78.235104} The estimated overall error of the method used is around 0.1 eV.

\section{Results}
\vspace {-3mm}
The acceptor levels of the carbon vacancy are found to show negative-$U$ and close to negative-$U$ property at the $k$- and the $h$-site, respectively. The difference between the (0/1-) and (0/2-) levels remains below 0.1 eV. We did not find any negative-$U$ behavior for the donor levels of the \VC. The difference between the (2+/1+) and the (1+/0) levels is around 0.1-0.2 eV. The positions of the acceptor and donor levels of \VC~occure around \EV+(2.7-2.8) eV and \EV+(1.6-1.8) eV, respectively, which can explain the levels of the \defEH~and \defZ~centers (see Table \ref{tab:occ}) (\EV~denotes the valence band maximum). The calculated formation energy of the neutral carbon vacancy is around 4.4 and 5.0 eV under Si-rich and C-rich condition, respectively. The cubic site is slightly more favorable than the hexagonal one. We also estimated the carbon vacancy concentration under typical growth conditions assuming that high temperature CVD growth occurs close to quasiequilibrium. Under such conditions, with a growth temperature of around 1600-1700 \celsius~and the $n$-doping concentration of around 3-5$\times$10$^\text{15}$ cm$^{-3}$, the approximated carbon vacancy concentration becomes $\sim$2$\times$10$^\text{12}$ cm${}^{-3}$ which is in the common concentration range of \defZ~and \defEH~centers in as-grown samples. \cite{APEX.2.091101}
\begin{table}[!b]
\setlength{\doublerulesep}{0 pt}
\setlength{\extrarowheight}{2 pt}
\begin{center}
\begin{tabularx}{34cc}{@{\hskip\tabcolsep\extracolsep\fill}l@{}cccccc@{}c} \toprule[1pt] \addlinespace[4pt]
\textbf{Defect}  & (2+/1+) & (1+/0) & (0/1-) & (1-/2-) & (2-/3-) & (2+/0) & (0/2-)\\
\addlinespace[4pt]
\midrule[0.50pt]
\addlinespace[4pt]
\VC($k$)           & 1.67       & 1.75    & 2.80   & 2.74      &             & 1.71    & 2.77 \\ \addlinespace[2pt]
\VC($h$)           & 1.64       & 1.84    & 2.71   & 2.79      &             & 1.74    & 2.75 \\ \addlinespace[4pt]
\VSi($k$)           &               &            & 1.24   & 2.47      & 2.74     &            & \\ \addlinespace[2pt]
\VSi($h$)           &               &            & 1.30   & 2.59      & 2.85     &            & \\ \addlinespace[3pt] \bottomrule[1pt]
\end{tabularx}
\end{center}
\caption{Calculated (HSE06) and corrected occupation levels of the vacancies. Values are in eV relative to the valence band maximum (E${}_\text{V}$).}
\label{tab:occ}
\end{table}
\begin{figure}[!t]
\centering
\includegraphics[scale = 0.70]{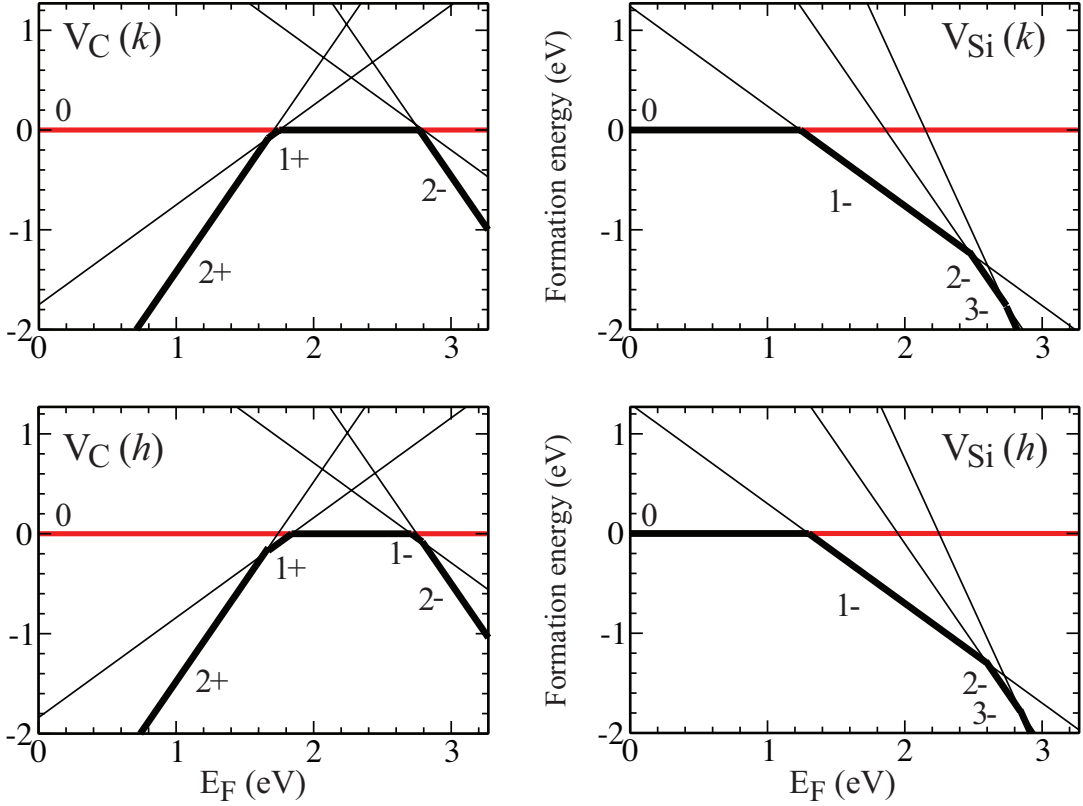}
\caption{Relative formation energy diagrams of the carbon vacancies (left column) and the silicon vacancies (right column). Occupation levels are at the intersection of the lines. The Fermi energy (E${}_\text{F}$) is relative to E${}_\text{V}$.}
\label{fig:occ}
\end{figure}

In contrast to standard DFT results, we did not find any donor levels for the silicon vacancy (\VSi). Further, previous results have suggested that \VSi~can exist even in the 4- charge state \cite{PhysRevB.59.15166}, while our results show that the 3- charge state can only exist in highly \textit{n}-type material and the (3-/4-) level is above the conduction band minimum. The calculated formation energy of the neutral silicon vacancy is around 8.0 and 7.5 eV under Si-rich and C-rich condition, respectively. Hence, for a wide range of doping concentrations the carbon vacancy is much more abundant than the silicon vacancy. Only under extreme C-rich conditions and high $n$-doping levels, the formation energy of \VSimmm~can approach that of \VC~(see Fig. \ref{fig:occ}).

\section{Discussion}
\vspace {-3mm}
The origin of the \defZ~and \defEH~centers has been disputed in the literature for a long time. The former is a negative-$U$ center with a sharp DLTS peak at around \EC-(0.6-0.7) eV, while the latter is broader peak with one or more contributions at around \EC-(1.5-1.6) eV (\EC~denotes the conduction band minimum). One of the possible microscopic origins for these centers is the carbon vacancy, and indeed, our calculated occupation levels of \VC~at \EC-(0.5-0.6) eV and at \EC-(1.5-1.7) eV are close to the experimentally observed level positions of the \defZ~and \defEH~defects, respectively. Although the negative-$U$ property of the acceptor level remains only in the case of the $k$-site, the occupation levels of the $h$-site are very close to each other, within 0.1 eV, and their negative-$U$ property disappears only after applying the charge correction. The donor levels of \VC~are also close to each other, within 0.2 eV, with no negative-$U$ behavior. This is consistent with the experimental findings that the \defEH~peak is broad with one or more contributions. The two close donor levels of  \VC~in the midgap can also explain the appearance of the \defEI~EPR signals, confirming the results by Umeda \textit{et al.} \cite{PhysRevB.70.235212} Nevertheless, it has to be noted that for an unambiguous identification of \defZ, \defEH~and \defEI~centers other microscopic candidates have to be investigated as well.

According to its formation energy, \VSi~is a much less abundant defect than \VC, and it exhibits only acceptor levels (without negative-$U$) in the band gap of \hsic. Its occupation levels may explain the acceptor level of the \VSi~proposed by Son \textit{et al.} \cite{PhysRevB.75.155204} in semi-insulating \hsic, with an activation energy around 0.8-0.9 eV, that is close to the calculated (1-/2-) level of the \VSi~defect at \EC-(0.7-0.8) eV. Furthermore, the calculated properties of \VSi~(formation energy and level positions) agree closely with those observed experimentally for the so-called S-centers \cite{david:4728} and further work is in progress to substantiate this identification experimentally.

In summary, with a highly accurate \textit{ab initio} calculations the positions of the occupation levels of the carbon and silicon vacancies in 4H-SiC have been estimated. The results suggest that the \defZ~and \defEH~centers originate from the carbon vacancy.

Financial support by the Norwegian Research Council (CAPSIC-project, FRINAT and NOTUR programs) and from the Hungarian OTKA Grant No. K-67886 and the J\'{a}nos Bolyai program from the Hungarian Academy of Sciences is gratefully acknowledged.

\end{document}